# Temperature dependence of irradiation hardening due to dislocation loops and precipitates in RPV steels and model alloys

S. Kotrechko[1], V. Dubinko[2,1], N. Stetsenko[1], D. Terentyev[3], Xinfu He[4]

[1]G.V. Kurdyumov Institute for Metal Physics of NAS of the Ukraine, Kiev, Ukraine,

[2]NSC Kharkov Institute of Physics and Technology, NAS of the Ukraine, Kharkov, Ukraine

[3]SCK•CEN, Nuclear Materials Science Institute, Boeretang 200, Mol, 2400, Belgium

[4]China Institute of Atomic Energy, 102413 Beijing, China

**Abstract**

A relative contribution to irradiation hardening caused by dislocation loops and solute-rich precipitates is established for RPV steels of WWER-440 and WWER-1000 reactors, based on TEM measurements and mechanical testing at *reactor operating temperature* of 563 K. The pinning strength factors evaluated for loops and precipitates are shown to be much lower than those obtained for model alloys based on the *room temperature* testing as well as those evaluated by means of atomistic simulations in the temperature range of 300 to 600 K. This discrepancy is explained in the framework of a model of thermally activated dislocation motion, which takes into account the difference in *temperature* and *strain rate* employed in atomistic simulations and in mechanical testing.

## 1. Introduction

Irradiation hardening and embrittlement of reactor pressure vessel (RPV) steels is one of the main factors that determine its operating life-time. The dependence of the yield stress on the irradiation dose strongly correlates with changes in the major microstructural features. So there have been a number of attempts to predict the yield stress changes in terms of physically-based structure-property relations [1 - 12]. It is generally admitted that the net hardening is mainly the result of two distinct mechanisms, namely: (i) precipitate-induced hardening and (ii) matrix hardening due to clusters of vacancies or self-interstitial atoms (SIAs) including dislocation loops (a.k.a. damage-induced hardening). The first mechanism is related to the radiation-enhanced diffusion of some solute elements and their clustering. It is known that one of the key elements responsible for the irradiation hardening and embrittlement of RPV steels is copper that results in the formation of so called copper-rich precipitates (CRP) [8 -13]. The second component of irradiation hardening is related to the formation of dislocation loops (DLs), some of which (with sizes lager than 2÷3 nm) are experimentally observed in irradiated RPV steels and RPV model alloys along with solute-rich precipitates [9-13]. According to the cluster dynamics modeling [14], nanosized DLs can evolve from displacement cascades under reactor irradiation at neutron fluences as low as $10^{22}$ neutron/m$^2$ (~ 0.001 dpa), i.e. long before their detection by transmission electron microscope (TEM). The model proposed in [14] predicts the evolution of the loop size distribution function with irradiation dose, which was used for the evaluation of the DL-induced hardening according to the dispersed barrier model (DBM) [15]

$$\Delta\sigma_Y = \alpha Gb\sqrt{N \times d}, \qquad (1)$$

where $N \times d$ is the product of the number density and diameter of the DLs, respectively, $G$ is the shear modulus and $b$ is the Burgers vector. An effective strength of a DL as an obstacle to the dislocation glide (a.k.a. pinning strength), α was assumed in ref. [14] to obey a purely heuristic dependence on the loop size ranging from 0.1 to 1. The predicted hardening was compared to experimentally observed values from ref. [11] simply neglecting the contribution from precipitates that were observed in [11]. A possible justification of this approximation can be found in the refs. [11, 12], where it was noted that the hardening increased monotonously with increasing irradiation dose in contrast to a rapid decrease of the precipitate number density, and it was concluded that the precipitate contribution to the irradiation hardening was insignificant after sufficiently high neutron fluences. However, for a more adequate comparison of the dislocation-induced hardening predicted by the model with experimental data, one should separate their contributions from the contribution due to precipitates.

---

[1] Corresponding author: vdubinko@mail.ru

Chaouadi et al [8] approached this problem by investigating the hardening of Fe alloys with different copper content assuming that the matrix hardening is independent of the precipitate-induced one. This assumption allowed the authors to separate the copper-rich precipitate and matrix damage components by using a linear superposition. However, the lack of microstructural characterization in [8] does not allow one to use the results for characterization of the pinning strength of each component. What is more, it should be noted that the above conclusion about independence of the two contributions was provided for the steels irradiated up to 0.04 dpa only, while, the observation of the well developed μ-structure was observed at 0.8 dpa. A detailed experimental analysis of the radiation-induced changes in the microstructure of WWER-440 and WWER-1000 RPV steels can be found in refs [9 - 12]. The precipitate contribution to the irradiation hardening was argued to be insignificant for WWER-440 RPV steels [9, 10], whereas for WWER-1000 RPV it was found to be essential [12]. An attempt to analyze the experimental trends in terms of the DBM (see Eq.1) has shown that the strength factors should strongly depend on the defect type, but the nature of this dependence was not rationalized for RPV steels so far.

In works [13, 16] a correlation was made between microstructural observations by various complementary techniques [13] and the hardening measurements of the 16MND5 RPV steel and several model alloys: Fe-Cu, Fe-Mn-Ni, Fe-Mn-Ni-Cu. In [15], the authors used the DBM and a phenomenological superposition law suggested by a parametric study. It was found that dislocation loops were more strong obstacles to dislocation motion as compared to the precipitates, but due to their low concentration, they played a minor role in the net hardening. These results have been obtained for comparatively low dose irradiation (<0.1 dpa), at which radiation defects are not yet observable in TEM, as compared to TEM results for RPV steels irradiated up to ~1 dpa.

In order to rationalize these observations one needs a tool for modeling the strength of the loops and precipitates of different size and composition. Molecular dynamics (MD) simulations is a natural way to study and evaluate the interaction of dislocations with nanometric defects and reveal the physical nature of the pinning as well as estimate the resulting pinning strength. Up to now, the interaction of edge and screw dislocations with interstitial DLs, vacancy clusters and Cu precipitates has been simulated in bcc iron [17-28]. It appears that the relative pinning strength of DLs and precipitates is strongly dependent on ambient temperature and defect size. At this, non coherent Cu precipitates with size > 4nm act as strong impenetrable obstacles, while smaller ones are shearable particles with the strength that quickly decreases with simulation temperature [20]. Voids act as strong obstacles already at small sizes i.e. above 2 nm (see e.g. [20, 21]), while DLs exhibit a dual nature and can be absorbed or bypassed depending on temperature and loop size (see e.g. [19, 21-23]). It should be noted, however, that for computational reasons, MD simulations are performed for dislocation velocities of $(2 \div 200)$ m·s$^{-1}$, which correspond to strain rates of $(10^6 \div 10^8)$ s$^{-1}$ that is about 10÷12 orders of magnitude higher the strain rate under typical mechanical testing (~$10^{-4}$ s$^{-1}$). To bridge the MD conditions and results with those of mechanical testing, the rate-controlling deformation mechanism must be understood. An integration of this complex dependence of the pinning strength of defects on temperature, strain rate, defect type and size into a single model is necessary to assess net hardening in the material exhibiting two or more hardening sources.

The goal of the present work is to make a comparative assessment of contributions of matrix damage, attributed to dislocation loops and precipitates to irradiation hardening of the of WWER-440 and WWER-1000 RPV steels over a wide range of neutron fluences. Consequently, in section 2, we summarize available experimental data regarding the radiation-induced microstructure and hardening at low and high fluences, and evaluate the strength factors of dislocation loops and precipitates contributing to hardening. The obtained set of the strength factors is compared with those obtained for model alloys [16] and with results of MD simulations reviewed in section 3. In section 4, effects due to temperature and strain rate on the pinning strength of defects are analyzed in the framework of a model of thermally activated dislocation motion, and the demonstrated discrepancy between different experimental data and MD simulations is explained. The results are discussed in Section 5 and summarized in Section 6.

## 2. Evaluation of DL and precipitate contribution to irradiation hardening

### 2.1. Power law approximation

Experimental data for irradiation hardening are usually characterized using a power law approximation [23, 24]:

$$\Delta\sigma_Y = B_h \left(\frac{\Phi}{\Phi_0}\right)^m, \qquad (2)$$

where $B_h$ is the coefficient and $m$ is the hardening exponent; $\Phi$ is the neutron fluence, $\Phi_0 = 10^{22}$ neutron/m$^2$. In case under consideration, the best approximation for BM and WM of WWER-1000 and for BM of WWER-440 (Table 1) was obtained at $m=0.5$ (Fig.1) and for WM of WWER-440 – at $m=0.33$, which agrees with typical $m$ values for RPV steel, which lie within the range of 0.33÷0.5 [23, 24].

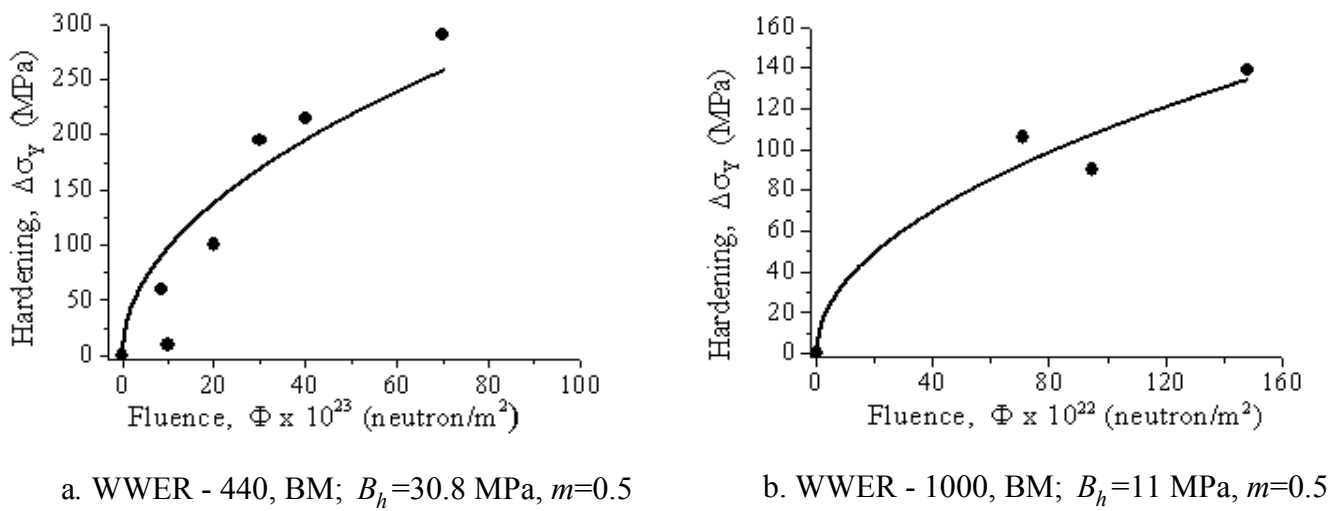

a. WWER - 440, BM; $B_h$=30.8 MPa, $m$=0.5      b. WWER - 1000, BM; $B_h$=11 MPa, $m$=0.5

Figure 1. Power law approximation (Eq. (5)) of experimental data (●) on irradiation hardening of RPV steels based on mechanical tests performed at 563 K [11, 12].

Table 1. Parameters of radiation-induced defects [11, 12]

| N | $\Phi$, $10^{23}$ n/m$^2$ | Density of DLs, $10^{21}$ m$^{-3}$ | Diameter of DLs, nm | Density of disk-shaped precipitates, $10^{21}$ m$^{-3}$ | Diameter of disk-shaped precipitates, nm | Density of rounded precipitates, $10^{21}$ m$^{-3}$ | Diameter of rounded precipitates, nm |
|---|---|---|---|---|---|---|---|
| \multicolumn{8}{c}{WWER-440, BM - 15Kh2MFA (P=0.012%, Cu= 0.11%)} |
| 1 | 10 | 9-10 | 5-7 | 5-6 | 17-20 | 800-900 | 2,5-3 |
| 2 | 30 | 20-30 | 7-9 | 5-6 | 16-18 | 1600-1800 | 3,5-4 |
| 3 | 66 | 30-40 | 3.5-4.5 | 6-8 | 16-18 | 1700-1900 | 2,5-3,5 |
| 4 | 80 | 50-70 | 8-10 | 4-5 | 15-17 | 800-1100 | 4-5 |
| 5 | 168 | 110-130 | 15-18 | 2-3 | 12-15 | 60-70 | 5-7 |
| \multicolumn{8}{c}{WWER -440, WM - Sv-10KhMFT (P=0.027%, Cu=0.04%)} |
| 6 | 8,6 | 8-9 | 4-6 | 20-30 | 17-19 | 700-800 | 2-3 |
| 7 | 30 | 30-40 | 5-6 | 30-40 | 15-17 | 2000-2500 | 2,5-3,0 |
| 8 | 60 | 50-70 | 7-8 | 40-50 | 14-16 | 1000-1500 | 3-4 |
| 9 | 189 | 90-120 | 12-15 | 60-80 | 10-12 | 50-70 | 5-6 |
| \multicolumn{8}{c}{WWER -1000, BM - 15Kh2NMFAA (Ni=1.34%, Mn= 0.47% Si=0.29%)} |
| 10 | 7,7 | 5-6 | 4-5 | 3-5 | 15-20 | 80-100 | 2-3 |
| 11 | 9,4 | 7-8 | 4-5 | 3-5 | 15-20 | 100-200 | 2-3 |
| 12 | 14,7 | 10-20 | 4-5 | 4-6 | 10-15 | 200-300 | 3-4 |
| \multicolumn{8}{c}{WWER -1000, WM - Sv-10KhGNMAA (Ni=1.77%, Mn= 0.74% Si=0.26%)} |
| 13 | 3,1 | 5-6 | 4-5 | - | - | 70-90 | 2-3 |
| 14 | 5,2 | 6-7 | 4-5 | - | - | 200-400 | 2-3 |
| 15 | 6,5 | 10-20 | 5-6 | - | - | 300-500 | 3-4 |
| 16 | 11,6 | 400-600 | 6-8 | - | - | 700-800 | 3-5 |

### 2.2. Validity of dispersed barrier model

The DBM is derived by considering a dislocation bowing between a square array of obstacles, and using an approximation of the *constant line tension*:

$$\text{T}(\theta) \approx \frac{Gb^2}{2}, \qquad (3)$$

which does not depend on the character (edge, screw or mix) of the dislocation segments [20]. More accurate expression has the following form [21 – 22]:

$$\text{T}(\theta) = \frac{Gb^2}{4\pi}\left(\frac{1+\nu-3\nu\sin^2\theta}{1-\nu}\right)\ln\left(\frac{L-2r}{2b}\right), \qquad (4)$$

where $2b$ is the dislocation core radius, $\nu$ is the Poisson ratio, $\theta$ is the angle between the segment line in its straight configuration and the Burgers vector, $L - 2r$ is the outer cut-off distance ($L$ is the distance between the centres of two obstacles pinning the segment and $r$ is the average radius of these obstacles in the slip plane). With account of (4), the factor $\alpha$ in eq. (1) can be written as

$$\Delta\sigma_Y = \alpha Gb\sqrt{N\times d}, \quad \alpha = \alpha_0 \cdot 0.85M\frac{F}{2\pi}\ln\left(\frac{1}{2b\sqrt{N\times d}}\right), \quad F \approx 1.25, \quad (5)$$

where $M = 2.75$ is Taylor factor for b.c.c. lattice, 0.85 is the statistical coefficient that accounts for random distribution of obstacles, $F$ characterises the type of dislocation, and it is presented for the case of mixed dislocations at $\nu \approx 0.33$. Here we have introduced a *pinning strength* factor $\alpha_0$, which should be equal to unity in the ideal case of impenetrable (hard) obstacles in the dispersed barrier model. However, the unpinning of dislocations from obstacles is a thermally activated process [20]. As a result, the pinning strength $\alpha_0$ may be less than unity, and it may strongly depend on temperature and strain rate, as will be demonstrated in Section 4.

In contrast to $\alpha_0$ the factor $\alpha$ given by eq. (5) depends not only on the obstacle type but also on their number density and mean size that evolve with irradiation time due to logarithmic dependence, which is usually neglected. However, as will be shown bellow (Table 3), the $\alpha$ value may change with neutron fluence by 70%, and neglecting the change imposes a large error on the hardening evaluation.

It is convenient to present the hardening due to the obstacles of a specific type "$i$", $\Delta\sigma_{Yi}$, as a product of a constant factor $\alpha_{0i} \cdot 0.85M\frac{F}{2\pi}Gb = 0.465\cdot\alpha_{0i}Gb$ and a structural function $f(N_i \times d_i)$, which depends only on the mean size and number density of specific defects and can be measured experimentally:

$$f(N_i \times d_i) = \ln\left(\frac{1}{2b\sqrt{N_i\times d_i}}\right)\sqrt{N_i\times d_i}, \quad (6)$$

Note that $L(N_i \times d_i) = 1/f(N_i \times d_i)$ is the effective distance between the obstacles. Structural functions for precipitates ($i = 1$) and DLs ($i = 2$), evaluated on the basis of experimental data for WWER-440 base metal (Table 1), are shown in Fig. 2. It can be seen that behaviour of structural function for DL agrees well with a typical experimental behaviour of irradiation hardening (see Fig. 1), which increases monotonically with neutron fluence, in contrast to a structural function for precipitates, which goes through the maximum and decreases after $4\times10^{24}$ neutron/m$^2$. Comparison of the Figs. 1 and 2 indicates clearly that DLs can be expected to dominate irradiation hardening at higher neutron fluences.

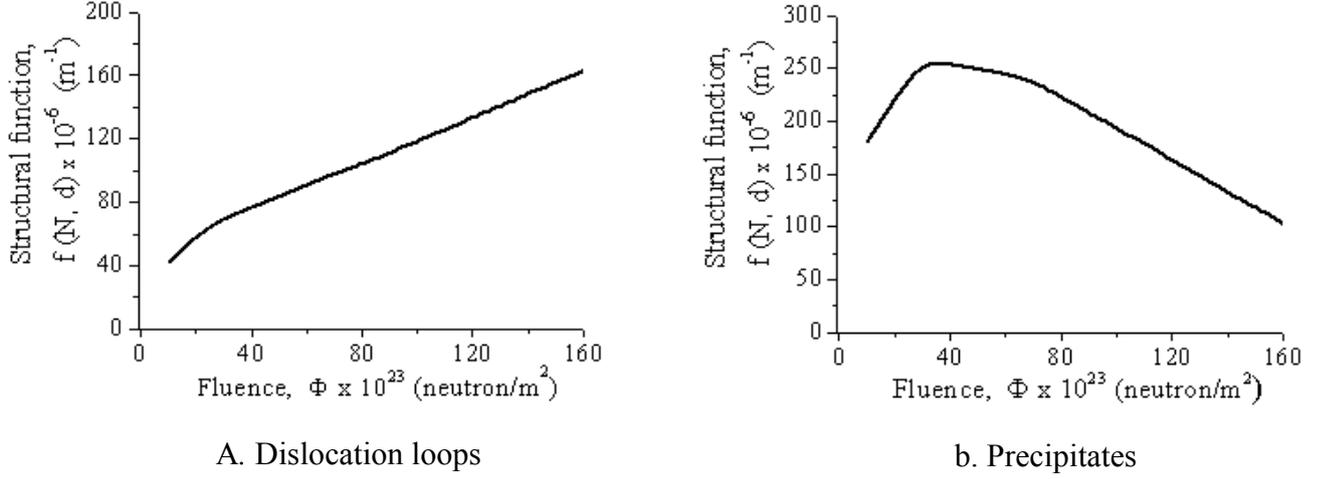

| A. Dislocation loops | b. Precipitates |

Figure 2. Structural functions for WWER-440, BM

*2.3. Selection of superposition law*

In general case, a superposition of two components of hardening, namely, of DL's and precipitates can be written as [8]:

$$\Delta\sigma_Y^P = \Delta\sigma_{Y1}^P + \Delta\sigma_{Y2}^P, \qquad (7)$$

where the upper script $P$ is the superposition parameter that depends on the obstacle strength, which is *a priori* unknown. Two extreme cases are the square-law superposition ($P=2$) used for the obstacles of comparable strength [15, 25] and the linear superposition ($P=1$) used for strong obstacles surrounded by a high concentration of weak obstacles [8]. An intermediate superposition law is considered if neither linear nor square-law superposition gives acceptable approximation of experimental data.

Now, in order to find the factors for the two main components $\alpha_{01}$ and $\alpha_{02}$, two experimental values of hardening at different fluences should be used, one – for small fluences, and the other – for high fluences. We will consider two opposite superposition laws using the approximating curves in Fig. 1 as experimental values in the calculations.

For the *square-law superposition* stress squares are additive:

$$\Delta\sigma_Y = \sqrt{\Delta\sigma_{Y1}^2 + \Delta\sigma_{Y2}^2}, \qquad (8)$$

which can be rewritten as follows with account of eqs. (5), (6):

$$\Delta\sigma_Y = 0.465 \cdot Gb\sqrt{\alpha_{01}^2 \cdot f^2(N_1 \times d_1) + \alpha_{02}^2 \cdot f^2(N_2 \times d_2)}, \qquad (9)$$

whence one obtains two equations for two variables $\alpha_{01}$ and $\alpha_{02}$ and their coefficients $f^2(N_1 \times d_1)$ and $f^2(N_2 \times d_2)$, which are determined at the same fluences as experimental values of hardening

$\Delta\sigma_{Y_I}^{exp.}$ and $\Delta\sigma_{Y_{II}}^{exp.}$:

$$\left(\frac{\Delta\sigma_{Y_I}^{exp.}}{0.465 \cdot Gb}\right)^2 = \alpha_{01}^2 \cdot \left[f^2(N_1 \times d_1)\right]_I + \alpha_{02}^2 \cdot \left[f^2(N_2 \times d_2)\right]_I, \quad (10)$$

$$\left(\frac{\Delta\sigma_{Y_{II}}^{exp.}}{0.465 \cdot Gb}\right)^2 = \alpha_{01}^2 \cdot \left[f^2(N_1 \times d_1)\right]_{II} + \alpha_{02}^2 \cdot \left[f^2(N_2 \times d_2)\right]_{II}, \quad (11)$$

For the *liner superposition* one has instead of eq. (8)-(11)

$$\Delta\sigma_Y = \Delta\sigma_{Y1} + \Delta\sigma_{Y2}, \quad (12)$$

$$\Delta\sigma_Y = 0.465 \cdot Gb \cdot \left[\alpha_{01} \cdot f(N_1 \times d_1) + \alpha_{02} \cdot f(N_2 \times d_2)\right], \quad (13)$$

$$\frac{\Delta\sigma_{Y_I}^{exp.}}{0.465 \cdot Gb} = \alpha_{01} \cdot \left[f(N_1 \times d_1)\right]_I + \alpha_{02} \cdot \left[f(N_2 \times d_2)\right]_I, \quad (14)$$

$$\frac{\Delta\sigma_{Y_{II}}^{exp.}}{0.465 \cdot Gb} = \alpha_{01} \cdot \left[f(N_1 \times d_1)\right]_{II} + \alpha_{02} \cdot \left[f(N_2 \times d_2)\right]_{II}, \quad (15)$$

Solution of equation sets (10) - (11) and (14) - (15) for experimental values of hardening, and defect parameters allows one to determine factors $\alpha_{01}$ and $\alpha_{02}$. It appears that $\alpha_{01} \ll \alpha_{02}$ for both superposition laws (Table 2), i.e. strength of more numerous precipitates is much lower than strength of DL's. This means that a linear superposition is more adequate in the case under consideration, and the resulting factors $\alpha_0$ are presented in Table 2 in comparison with factors $\alpha$. It can be seen that $\alpha$ change with dose significantly due to the microstructure evolution in contrast to $\alpha_0$ that are much more stable and reflect the corresponding defect strengths rather than their evolution.

Table 2. Factors $\alpha_0$ vs. $\alpha$ evaluated in the fluence range $10^{24}$-$10^{25}$ n m$^{-2}$

| Material | Precipitates, linear superposition | | | DLs, linear superposition | | |
|---|---|---|---|---|---|---|
| | $d_{mean}$ (nm) | $\alpha_1$ | $\alpha_{01}$ | $d_{mean}$ (nm) | $\alpha_2$ | $\alpha_{02}$ |
| BM, WWER-440 | 2.7÷6 | 0.03÷0.043 | 0.02 | 6÷17 | 0.47÷0,32 | 0.18 |
| WM, WWER-440 | 2.5÷5.5 | 0.038÷0.052 | 0.024 | 5÷14 | 0.39÷0.27 | 0.148 |
| BM, WWER-1000 | 2.5÷3.5 | 0.085÷0.073 | 0.04 | 4÷5 | 0.52÷0.48 | 0.2 |
| WM, WWER-1000 | 2.5÷4 | 0.046÷0.033 | 0.02 | 5÷7 | 0.56÷0.34 | 0.2 |

Irradiation hardening due to radiation-induced DLs and precipitates evaluated using the evaluated factors $\alpha_{0i}$ are shown in Fig. 3.

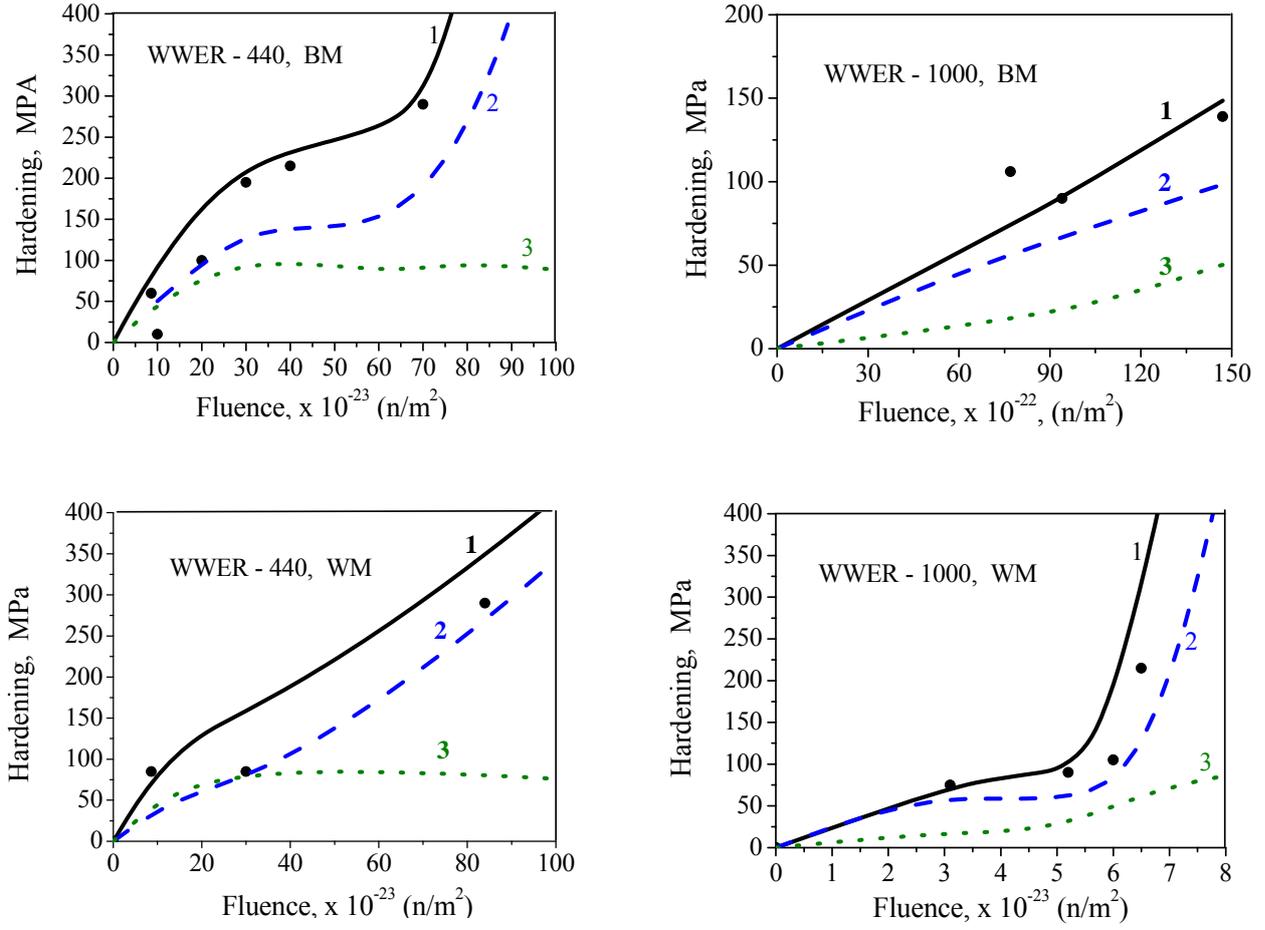

Figure 3. Contributions of DL's and precipitates to irradiation hardening of RPV steels (linear superposition based on mechanical tests performed at 563 K): ● - experimental data [11, 12]; the curve **1** – $\Delta\sigma_Y$ by eq. (8); the curve **2** – hardening due to DLs, **3** – hardening due to precipitates.

One can see that in all cases but one (WWER - 1000, BM), precipitate contribution to hardening becomes insignificant at high doses due to severe decrease in their number density caused by the precipitate coarsening. At lower dose levels, the contributions to hardening from DLs and precipitates are comparable with slight domination of DLs. This is a consequence of that the pinning strength factors of precipitates are very small as compared with DL, which agrees *qualitatively* with conclusions made in [10-12, 16, 29], but disagrees *quantitatively* with them as well as with molecular dynamics (MD) simulations of the interaction of dislocations with DLs and precipitates [17-28] (see the following section 3). However, a quantitative comparison of the presented $\alpha_0$ values with those deduced from MD studies must be done only on the basis of comprehensive theory developed in Section 4.

## 3. MD simulations of the dislocation interaction with DLs and precipitates

MD simulations is a natural way to study and evaluate the interaction of dislocations with nanometric defects and reveal the physical nature of the pinning as well as estimate the resulting pinning strength, which is evaluated as follows.

The computer based elasticity treatment of dislocation self-stress during its interaction with a regular array of impenetrable obstacles (Orowan strengthening) or voids results in the following expression for the critical stress required for the dislocation unpinning [1, 20, 30]

$$\tau_c = \frac{Gb}{2\pi L}\left[\ln\left(D^{-1}+L^{-1}\right)+\Delta\right], \qquad (16)$$

where $D$ is the obstacle size, $L+D$ is the spacing between the obstacle centers, and $\Delta$ depends on the obstacle type ranging from 0.77 to 1.52 (Fig. 4a).

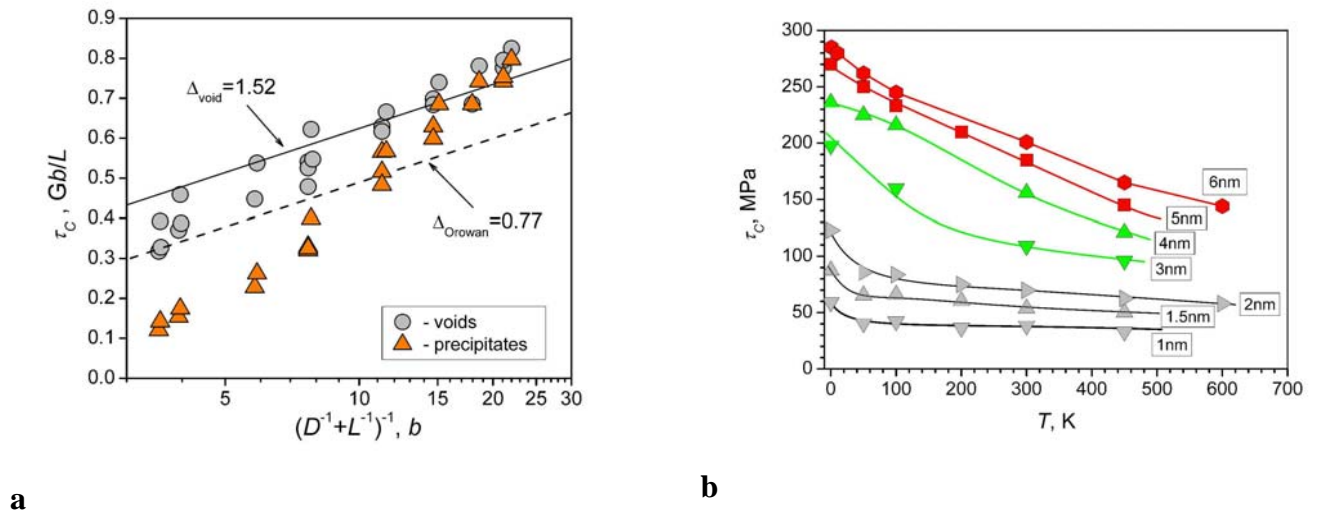

a  b

Figure 4. (a) Critical stress $\tau_c$ versus the harmonic mean of D and L for voids (circles) and Cu precipitates (triangles) in Fe at 0 K for the values of D (0.9÷6 nm) and $L+D$ (41.4÷83.6 nm) MD modeled in [20]. Lines correspond to be best fits to $\tau_c$ values obtained in continuum modeling for impenetrable Orowan particles and voids [1, 30]. (b) Critical stress $\tau_c$ versus temperature obtained for precipitates of different sizes at $L+D$ = 41.4 nm and strain rate $\dot{\varepsilon}_{MD}$ = 5×10$^6$ s$^{-1}$ [20].

Critical stress obtained by MD modeling may be lower or higher than that given by eq. (16) depending on the obstacle size (Fig. 4a) and what is more, it *decreases with temperature* (Fig.4b) pointing out at the thermally-activated nature of the unpinning process. A *pinning strength* factor $\alpha_0$ can be introduced in eq. (16) similar to the eq. (5)

$$\tau_c = \alpha_0 \frac{Gb}{2\pi L}\left[\ln\left(D^{-1}+L^{-1}\right)+\Delta\right], \qquad (17)$$

whence it can be evaluated from the critical resolved shear stress obtained by MD as follows:

$$\alpha_0 = \frac{\tau_c L}{Gb} \frac{2\pi}{\left[\ln\left(D^{-1} + L^{-1}\right) + \Delta\right]} \tag{18}$$

The resulting $\alpha_0$ for DLs and precipitates are shown in Fig. 5. Data located in a grey box entitled 'carbon decoration' denote the spread of strength of 1nm loops decorated by carbon atoms as was studied in [26]. Data located in a grey box entitled 'screw dislocation' denote the results obtained for the interaction with a screw dislocation [23]. Data in orange box correspond to the results obtained for the $a_0$<100> loops interacting with edge dislocation. All other data is for the interaction of a0/2<111> loops with a0/2<111>{110} edge dislocation. For all loops, it is assumed that $\Delta = 1.52$ since they belong to the matrix damage similar to voids, in contrast to precipitates, for which it is assumed that $\Delta = 0.77$.

MD results indicate that the pinning strength increases with the defect size and decreases with temperature, as may be expected. However, the MD values of $\alpha_0$ greatly exceed those obtained from the present evaluation of the pinning strength of DLs with D > 5 nm ($\alpha_0 \sim 0.2$) and precipitates with D > 2.5 nm ($\alpha_0 \sim 0.02$) on the basis of the yield stress measured at 563 K (Table 2). In order to explain this discrepancy and to correlate the MD results with experimental data, let us consider effects due to the difference in temperature and strain rate in MD simulations and mechanical testing.

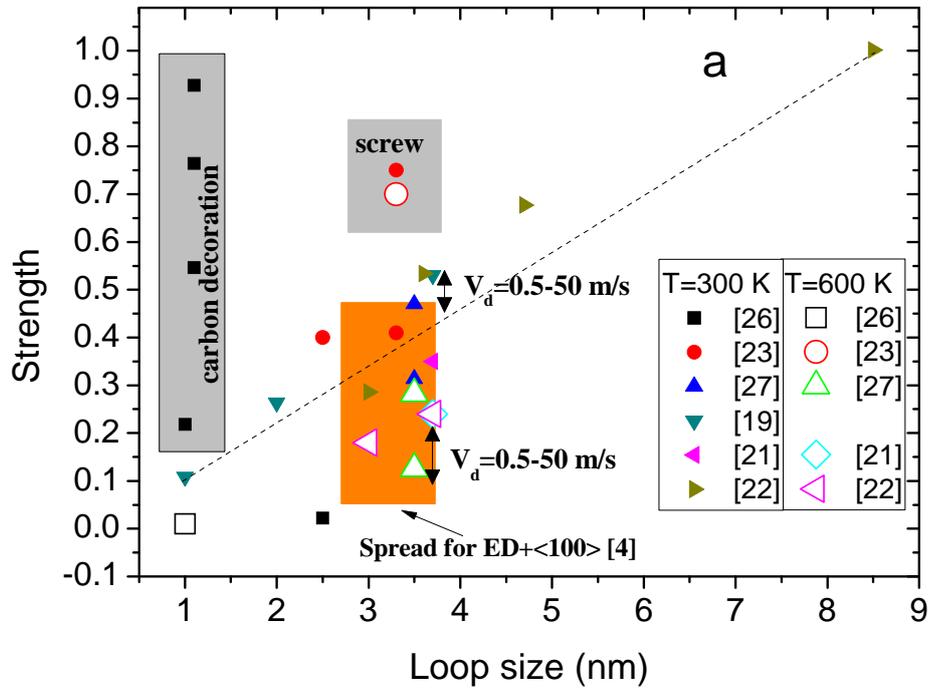

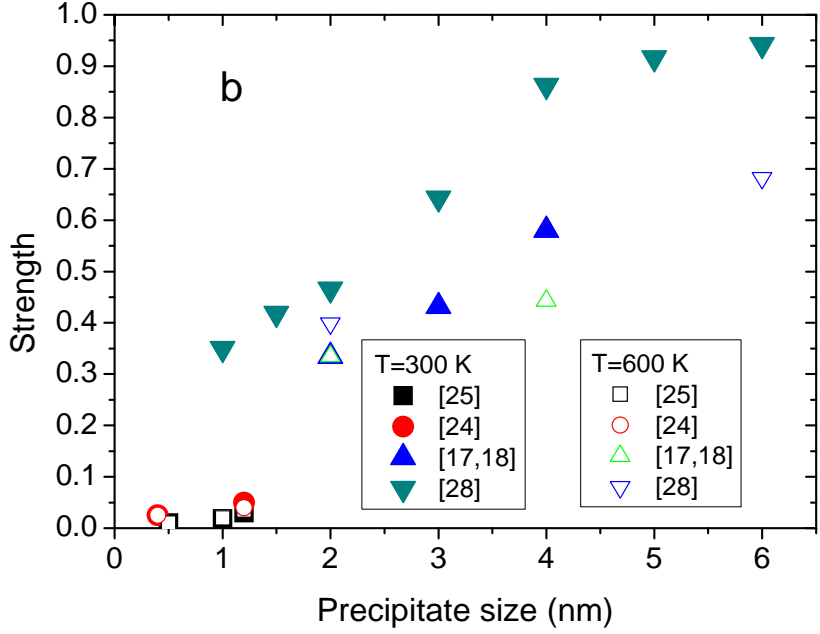

Figure 5. (a) Pinning strength $\alpha_0$ of DLs ($\Delta$ = 1.52); (b) $\alpha_0$ for precipitates ($\Delta$ = 0.77) at room temperature and 600 K obtained by MD simulations at strain rate $\dot{\varepsilon}_{MD}$ =10$^7$ s$^{-1}$ [17-19, 21-28].

## 4. Effect of temperature and strain rate on the pinning strength of defects

To describe the effect of strain rate $\dot{\varepsilon}$ on the yield stress $\sigma_Y$, it is conventionally written according to the Zerilli–Armstrong equation [31]

$$\sigma_Y = \sigma_a + \sigma_{th}(0)\exp\{-\beta T\}, \quad \beta = C_2 - C_3 \ln \dot{\varepsilon} \tag{19}$$

where $T$ is the temperature of mechanical testing, $\sigma_a$ is the athermal component of the yield stress, which measures the elastic limit of material without the aid of thermal fluctuations. It is determined mainly by long-ranged stress fields from large microstructural features, while the second term is due to small defects, from which dislocations can unpin by thermal fluctuations when the applied load is below the athermal limit $\sigma_{th}(0)$. The physical nature of other coefficients can be explained based on the following expression for the strain rate that is valid for bcc and fcc metals under sufficiently low temperatures [32]

$$\dot{\varepsilon} = \dot{\varepsilon}_0 \exp\left\{-\frac{U_0}{k_B T} \ln \frac{\sigma_0 - \sigma_a}{\sigma - \sigma_a}\right\}, \quad \dot{\varepsilon}_0 \approx b v_d^0 \rho_d, \tag{20}$$

where $\sigma_0$ is the yield stress at zero K, $U_0$ is the effective activation energy, $k_B$ is the Boltzmann constant, and $\rho_d$ is the density of dislocations that would glide at a rate $v_d^0$ in the limit of zero $U_0$.

Solving eq. (20) with respect to stress $\sigma$, one obtains Zerilli–Armstrong equation with the coefficients that are related to the microstructure parameters as follows:

$$\sigma_{th}(0) = (\sigma_0 - \sigma_a), \; C_2 = \frac{k \ln \dot{\varepsilon}_0}{U_0}, \; C_3 = \frac{k}{U_0} \Rightarrow \beta = \frac{k \ln(\dot{\varepsilon}_0/\dot{\varepsilon})}{U_0} \qquad (21)$$

Typical values of the coefficients for the 15Kh2MFA steel in unirradiated state are as follows [33] $\sigma_a$ =400 MPa, $\sigma_{th}(0)$=1526 MPa, $\beta$ = 7.771x10$^{-3}$ K$^{-1}$. The component $\sigma_{th}(0)$ correspond to the Orowan stress for small defects, which is given by the DBM similar to eq. (5)

$$\sigma_{th}(0) = \alpha_{th}(0) Gb\sqrt{N_{th} \times d_{th}}, \; \alpha_{th}(0) = \alpha_0^{th} 0.85 M \frac{F}{2\pi} \ln\left(\frac{1}{2b\sqrt{N_{th} \times d_{th}}}\right) \qquad (22)$$

where $d_{th}$ and $N_{th}$ are their size and number density, respectively. Assuming that $d_{th} \sim$ b and $\alpha_{0th}$ <0.1 (see Fig 4a) one obtains the following estimate for their concentration $N_{th}\omega \sim 1$ at% ($\omega$ is the atomic volume) which correlates with the concentration of impurity atoms in the material. Based on the known value of $\beta$, $\dot{\varepsilon}$ = 4×10$^{-4}$ s$^{-1}$, $\rho_d$ =10$^{14}$ m$^{-2}$, eq. (21) gives the following estimate for the underlying activation energy of the unpinning from impurities: $U_0$ = 0.286 eV. Their contribution to the yield stress dominates at sufficiently low temperatures (<RT), while it becomes insignificant at the reactor operating temperatures (~600 K).

Taking the difference between the yield stress before and after irradiation, one obtains from eq. (19) the following expression for the irradiation hardening

$$\Delta\sigma_Y(T) = \Delta\sigma_a + \sigma_{th}^0(0)\exp\{-\beta^0 T\} + \Delta\sigma_{th}^{irr}(0)\exp\{-\beta^{irr}T\} - \sigma_{th}^0(0)\exp\{-\beta^0 T\}, \qquad (23)$$

where the change of athermal component $\Delta\sigma_a = \sigma_a^{irr} - \sigma_a^0$ is expected to be insignificant, since it is determined mainly by long-ranged stress fields from large microstructural features, which do not change significantly, at least at the relevant irradiation doses. The second term in eq. (23) corresponds to initial solute obstacles, which remain essentially unperturbed by irradiation, and so it is canceled out with the forth term. The main contribution to irradiation hardening comes from radiation-induced nanometric DLs or precipitates, which is described by the third term in eq. (23). The activation energy of the unpinning from radiation-induced nanometric defects is expected to be higher than that for atomic size impurity defects present before irradiation, as will be verified bellow. For these reasons, eq. (23) can be rewritten in the following form:

$$\Delta\sigma_Y(T) \approx \Delta\sigma_{th}^{irr}(0)\exp\{-\beta^{irr}T\}, \; \beta^{irr} = \frac{k\ln(\dot{\varepsilon}_0/\dot{\varepsilon})}{U_0^{irr}}, \qquad (24)$$

where $U_0^{irr}$ is the effective activation energy of unpinning from radiation-induced defects, and the yield stress at zero K, $\sigma_{th}^{irr}(0)$, can be evaluated by the DBM similar to eq. (22)

$$\Delta\sigma_{th}^{irr}(0) = \alpha_{irr}(0) Gb\sqrt{N_{irr} \times d_{irr}}, \quad \alpha_{irr}(0) = \alpha_0^{irr} 0.85 M \frac{F}{2\pi} \ln\left(\frac{1}{2b\sqrt{N_{irr} \times d_{irr}}}\right), \quad (25)$$

where $d_{irr}$ and $N_{irr}$ are the mean size and number density of radiation-induced defects, respectively. For different types and sizes of defects, one can write, based on eqs (24) and (25), the following linear approximation for the irradiation hardening

$$\Delta\sigma_Y(T) \approx \sum_i \sigma_{th,i}^{irr}(0)\exp\{-\beta_i^{irr}T\}, \quad \beta_i^{irr} = \frac{k\ln(\dot\varepsilon_0/\dot\varepsilon)}{U_{0,i}^{irr}}, \quad \sigma_{th,i}^{irr}(0) = \alpha_{irr,i}(0) Gb\sqrt{N_{irr,i} \times d_{irr,i}} \quad (26)$$

where "$i$" is the type/size of the matrix damage or precipitate affecting the dislocation mobility. It is convenient to rewrite eq. (26) in the following form:

$$\Delta\sigma_Y(T) \approx \sum_i \sigma_{th,i}^{irr}(T), \sigma_{th,i}^{irr}(T) = \alpha_{irr,i}(T) Gb\sqrt{N_{irr,i} \times d_{irr,i}}, \quad (27)$$

$$\alpha_{irr,i}(T) = 0.85 M \frac{F}{2\pi} \ln\left(\frac{1}{2b\sqrt{N_{irr} \times d_{irr}}}\right)\alpha_{0,i}^{irr}(T), \quad \alpha_{0,i}^{irr}(T) = \alpha_{0,i}^{irr}(0)\exp\{-\beta_i^{irr}T\} \quad (28)$$

Eq. (28) unravels the thermally-activated nature of the pinning strength of radiation-induced defects $\alpha_{0,i}^{irr}(T)$ entering eq. (5), and describes its dependence on temperature and strain rate explicitly. This dependence can not be directly investigated by means of MD, since for computational reasons, MD simulations are performed for dislocation velocities of $(2 \div 200)$ m·s$^{-1}$, which correspond to strain rates of $(10^6 \div 10^8)$ s$^{-1}$ that is about $10 \div 12$ orders of magnitude higher the strain rate under typical mechanical testing ($\sim 10^{-4}$ s$^{-1}$).

To bridge the MD conditions and results with those of mechanical testing, the rate-controlling deformation mechanism must be understood [34, 35]. In the model of thermally activated dislocation motion, its velocity and the corresponding strain rate is given by [34]

$$v_d = v_d^0 \exp\left(-\frac{U_0 - V(\tau_c - \tau_0)}{k_B T}\right), \quad \dot\varepsilon = b\rho_d v_d = \dot\varepsilon_0 \exp\left(-\frac{U_0 - V(\tau_c - \tau_0)}{k_B T}\right), \quad \dot\varepsilon_0 = b v_d^0 \rho_d, \quad (29)$$

where $U_0$ is the activation energy, $\tau_c$ is critical applied shear stress, $\tau_0$ is the Peierls stress and $V$ is the activation volume. This equation is similar (but not equivalent!) to the phenomenological eq. (20), and it can be used in order to evaluate the unpinning activation energy $U_0$ that determines temperature dependence of the pinning strength $\alpha_{0,i}^{irr}(T)$ according to eq. (28). From eq. (29), it follows that $U_0$ can be expressed via the critical stress $\tau_c$ obtained by MD simulation at a constant strain rate $\dot{\varepsilon}_{MD}$ and temperature $T_{MD}$:

$$U_0 = V(\tau_c - \tau_0) + k_B T_{MD} \ln\left(\frac{\dot{\varepsilon}_0}{\dot{\varepsilon}_{MD}}\right), \tag{30}$$

For a quantitative evaluation of $U_0$ one needs to know the activation volume, which can be obtained e.g. by fitting the curve of strain rate as a function of yield stress, as has been demonstrated for a screw dislocation motion in iron [35]. This problem will require a lot of computational efforts to model defects of different sizes, which is beyond the scope of the present paper. Instead, we will use a rough approximation for the activation volume given by

$$V(D) \approx b^2 D, \tag{31}$$

that reflects our assumption of $V$ being proportional to the size of the obstacle, at which the applied force is concentrated. For example, $V$ (5nm) = 20 $b^3$, which is close to the value obtained by MD in single crystal iron for a typical stress [34].

Substituting eq. (30) into (28), one connects $\alpha_{0,i}^{irr}(T)$ to its value at 0 K, $\alpha_{0,i}^{irr}(0)$, that is determined by the critical stress at 0 K (eq. (18)), which can be evaluated from MD results obtained at $T_{MD}$ by the following equation:

$$\tau_c(0) = \tau_c(T_{MD}) + \frac{T_{MD}}{V} \ln\left(\frac{\dot{\varepsilon}_0}{\dot{\varepsilon}_{MD}}\right) \tag{32}$$

The results of such evaluation based on MD simulations [17-19, 21-28] and material parameters (Table 3) are presented in Table 4 and in Fig. 6. It can be seen that the pinning strength of nanometric defects crucially depends on temperature, which makes small precipitates and DLs (<3nm) extremely weak obstacles at reactor operating temperatures (563 K). The pinning strength of larger defects agrees well with phenomenological values from Table 2, as shown in Fig. 6.

For mechanical testing at ambient temperature (300K), the model predicts considerably higher pinning strength values, which agree well with phenomenological values obtained for model alloys at 300 K [16]: Fe-Cu, Fe-Mn-Ni, Fe-Mn-Ni-Cu, which are also shown in Fig. 6 for comparison (note that $\alpha_0 \sim 0.5\ \alpha$ evaluated in [16]).

On the other side, temperature dependence of the pinning strength of precipitates of different sizes calculated by eq. (28) at $\dot{\varepsilon}_{MD} = 10^7$ s$^{-1}$ is rather weak, in a qualitative agreement with direct MD results shown in Fig. 7, which verifies the assumptions made in evaluating the unpinning activation energy and volume.

We may conclude that significant difference in the pinning strength presented above for different simulation and experimental conditions is explained in the framework of a model of thermally activated dislocation motion, which takes into account the difference in *temperature* and *strain rate* employed in atomistic simulations and in mechanical testing.

Table 3. Material parameters for used in the present calculations and MD simulations

| *Parameter* | *Value* |
|---|---|
| Lattice constant, $b$, cm | $2.48 \times 10^{-8}$ |
| Shear modulus of RPV steel, $G$, GPa | 83 |
| Shear modulus of iron in MD simulations, $G_{MD}$, GPa | 83 |
| Dislocation density in RPV steel, $\rho_d$, m$^{-2}$ | $10^{14}$ |
| Dislocation density in MD simulations for DLs, $\rho_{MD}$, m$^{-2}$ | $1.67 \times 10^{15}$ |
| Dislocation density in MD simulations for CRP, $\rho_{MD}$, m$^{-2}$ | $8.3 \times 10^{14}$ |
| Strain rate in mechanical testing, $\dot{\varepsilon}$, s$^{-1}$ | $4 \times 10^{-4}$ |
| Strain rate in MD simulations, $\dot{\varepsilon}_{MD}$, s$^{-1}$ | $10^7$ |
| Peierls stress in MD simulations, $\tau_0$, MPa | 24 |
| Poisson ratio, $\nu$ | 0.33 |
| Taylor factor, $M$ | 2.75 |
| Temperature at mechanical testing of RPV, T, Kelvin | 563 |
| Temperature in MD simulations, $T_{MD}$, K, Kelvin | 300 |
| Free glide dislocation velocity, $v_d^0$, m·s$^{-1}$ | $10^3$ |

Table 4. Pinning strength of precipitates and DLs vs. size and temperature at a strain rate $4 \times 10^{-4}$ s$^{-1}$.

| D (nm) | | Precipitates | | | DLs $\frac{1}{2}<111>$ | | |
|---|---|---|---|---|---|---|---|
| Precipitates | DLs $\frac{1}{2}<111>$ | $U_0$ (eV) | $\alpha_{01}$ (300K) | $\alpha_{01}$ (563K) | $U_0$ (eV) | $\alpha_{02}$ (300K) | $\alpha_{02}$ (563K) |
| 3 | 3 | 0.199 | 0.041 | 0.002 | 0.193 | 0.02 | 0.0001 |
| 4 | 3.6 | 0.311 | 0.144 | 0.022 | 0.325 | 0.104 | 0.017 |
| 5 | 4.7 | 0.408 | 0.232 | 0.056 | 0.508 | 0.234 | 0.074 |
| 6 | 8.5 | 0.509 | 0.312 | 0.1 | 1.522 | 0.727 | 0.496 |

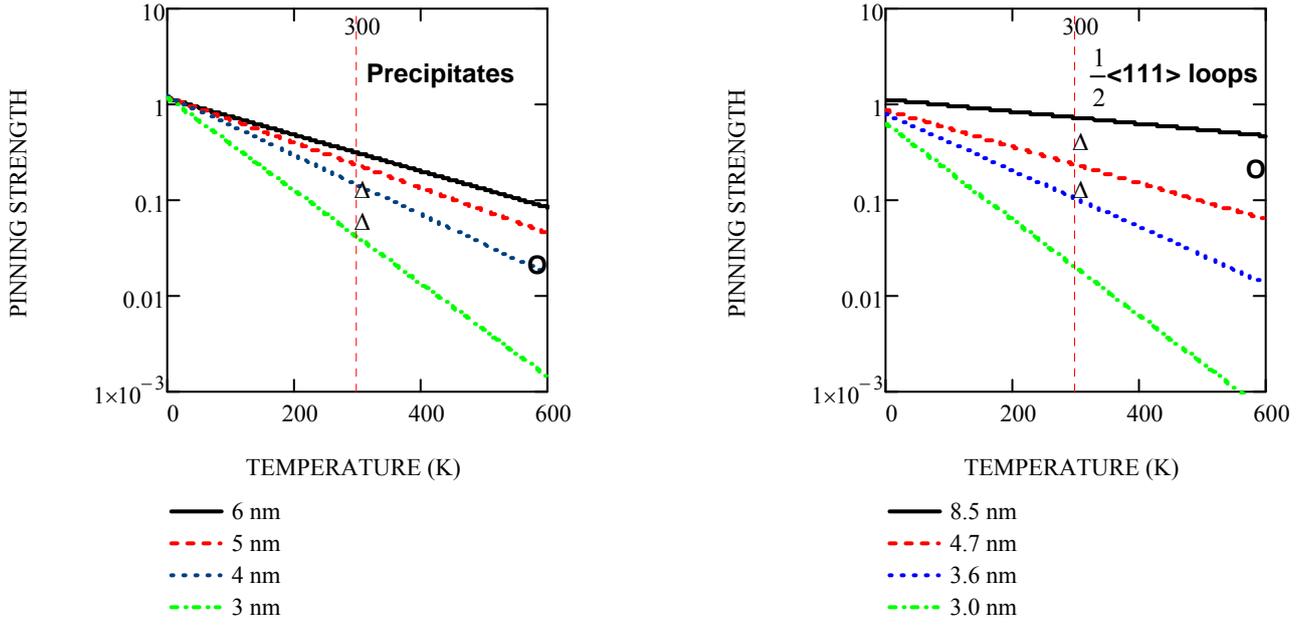

Figure 6. Temperature dependence of the pinning strength of precipitates and $\frac{1}{2}<111>$ loops of different sizes calculated by eq. (28) at $\dot{\varepsilon} = 4\times10^{-4}$ s$^{-1}$ using activation energy from Table 4 and material parameters from Table 3. Symbols **O** correspond to RPV experimental data (Table 2) obtained at **563 K** for loops with sizes ~ 6 nm ($\alpha_0$ ~0.2) and precipitates with sizes ~3 nm ($\alpha_0$ ~0.02). Symbols Δ correspond to experimental data in model alloys [16] obtained at **300 K** for loops with sizes ~ 4÷8 nm ($\alpha_0$ ~0.15÷0.35) and precipitates with sizes ~2÷4 nm ($\alpha_0$ ~0.06÷0.12).

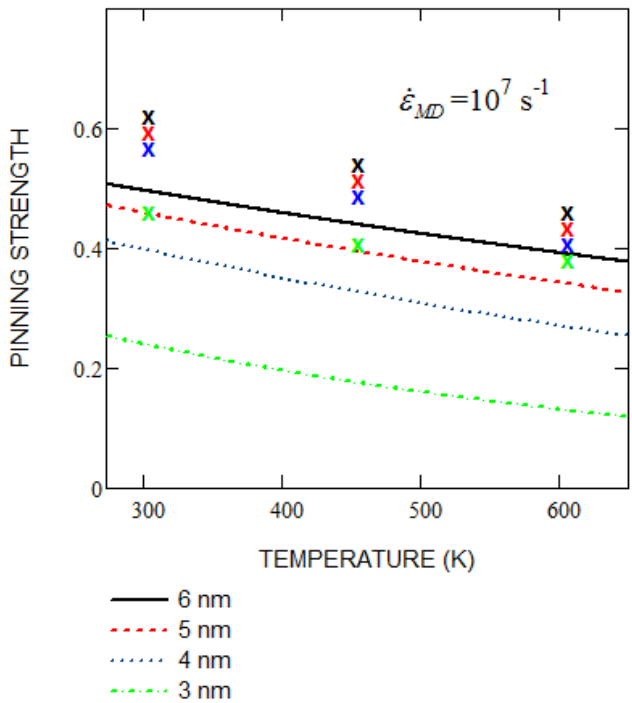

Picture 7. Temperature dependence of the pinning strength of precipitates of different sizes calculated by eq. (28) at $\dot{\varepsilon}_{MD} = 10^7$ s$^{-1}$, L= 41.4 nm, using activation energy from Table 4 and material parameters from Table 3. Symbols **X** correspond to the critical stress values from Fig. 4b and eq. (18).

## 5. Discussion

In the present paper, we have demonstrated that main reason for the apparent discrepancy between MD-based and experimental evaluation of the pinning strength of obstacles, is due to the huge difference in *the strain rate* employed in atomistic simulations and in mechanical testing. It is worth mentioning that another reason of the discrepancy may be based on a statistical nature of the unpinning from obstacles randomly arranged in crystal, which imposes dispersion in their spacing. Consequently, the critical stress $\tau_c$ is determined by the maximum obstacle spacing $L = L_{max}$ rather than by the average one, which is known as a concept of the "weak link" [36]. According to it, the event probability $P(\tau > \tau_c)$ is given by

$$P(\tau > \tau_c) = 1 - \exp\left[-N\left(\frac{\tau - \tau_{th}}{\tau_u}\right)^m\right], \qquad (33)$$

where $\tau_u$ and $m$ are the scale and shape parameters of the Weibull distribution, $\tau_{th}$ is the minimum Orowan stress for $L = L_{max}$, and $N$ is the number of obstacles along the dislocation front of the length $L_d$. At typical values of $L \approx 10 \div 100$ nm and $L_d \approx 10^4$ nm, one has $N \sim 10^2 \div 10^3$, and consequently, the mean critical stress approaches the minimum value:

$$\bar{\tau}_c \to \tau_{th} = \frac{\alpha G b}{L_{max}}, \qquad (34)$$

which may be lower than the theoretical value by a factor that depends on the dispersion of defect spacing. This example shows that a quantitative comparison between the theory and experiment is not straightforward and may require more detailed information about the real microstructure under investigation.

In spite of statistical uncertainties of the present results mentioned above, they allow one to reconsider some typical misbelieves concerning the temperature independence of irradiation hardening, expressed e.g. in [33, 37]: "The matrix damage and element precipitation result in an increase of $\sigma_Y$ as the lattice defects and precipitates affect the dislocation mobility. This increase of the yield stress is caused by an increase of the temperature-independent (athermal) component of the yield stress". Such a conclusion has resulted from the examination of irradiation hardening of RPV steel based on mechanical testing in the temperature range of $0 \div 200$ K [33], in which the yield stress is determined by impurities that have extremely high concentration (~1 at%) and relatively low pinning strength ($U_0 = 0.28$ eV) as compared to radiation-induced nanometric defects (see eq. (22) and below). For that reason, the radiation-induced contribution to the yield stress measured below room temperature is insignificant, while it becomes dominant with increasing temperature above RT up to the reactor operating temperatures (~600 K), at which small impurity obstacles become negligibly weak. While the pinning strength of radiation-induced nanometric defects is a strong function of temperature, as has been shown in the present paper, this should result in the temperature dependence of irradiation hardening. This conclusion has been supported by experimental evidence based on mechanical testing of RPV steels in the temperature range of 20÷350 ºC [38] demonstrated in Fig. 8.

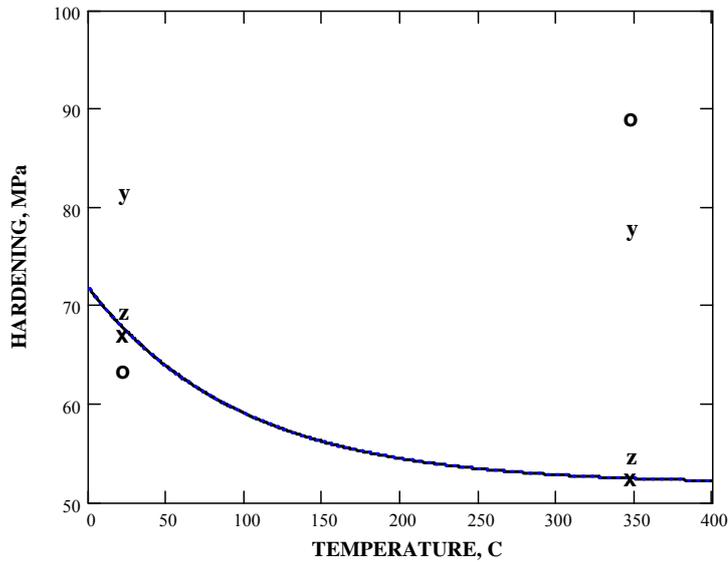

Figure 8. Dependence of irradiation hardening $\Delta\sigma_Y$ ($T$) of 15KhNMFA steel irradiated in WWER-1000 on the temperature of mechanical testing [38]. Symbols correspond to the following neutron fluences: **O** (20ºC) – $1.6\times10^{23}\,n\cdot m^{-2}$; **X** (20ºC) – $1.4\times10^{23}\,n\cdot m^{-2}$; **Z** (20ºC) – $2.5\times10^{23}\,n\cdot m^{-2}$; **y** (20ºC) – $1.8\times10^{23}\,n\cdot m^{-2}$; **O** (350ºC) – $2.1\times10^{23}\,n\cdot m^{-2}$; **X** (350ºC) – $2.3\times10^{23}\,n\cdot m^{-2}$; **Z** (350ºC) – $2.1\times10^{23}\,n\cdot m^{-2}$; **y** (350ºC) – $1.8\times10^{23}\,n\cdot m^{-2}$. The curve corresponds to $\Delta\sigma_Y$ ($T$) according to the model [38] for the neutron fluence of $2.5\times10^{23}\,n\cdot m^{-2}$.

One can see that a majority of experimental data points out at decreasing $\Delta\sigma_Y$ with increasing temperature of mechanical testing even for increased irradiation dose (the only exception corresponds to abnormally low initial yield stress before irradiation [38]). Similar trend is clearly visible in the extensive compilation of data on irradiation hardening in Fe-(8-9%Cr) steels irradiated in a wide dose range of 0.1÷94 dpa [39] (Fig. 9), which gives a strong support to the present results.

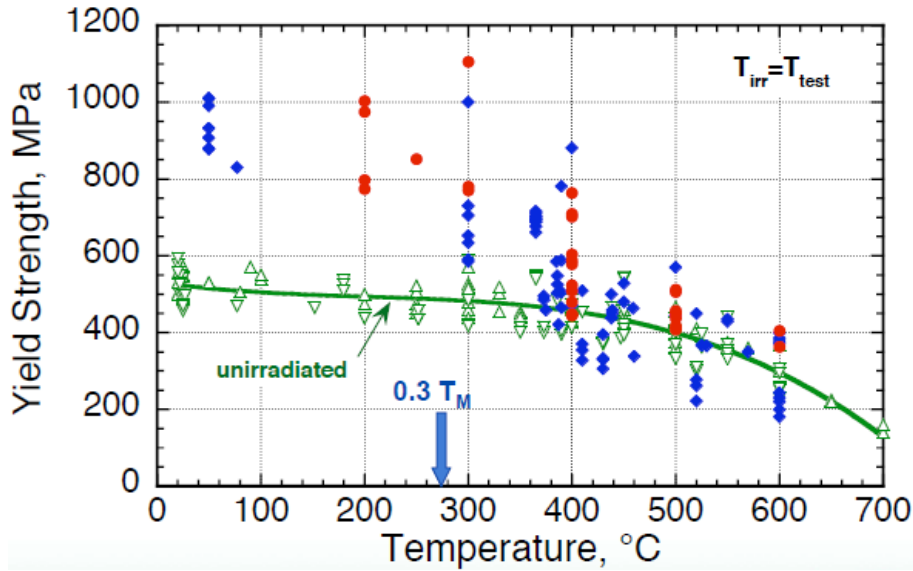

Figure 9. Irradiation hardening in Fe-(8-9%Cr) steels irradiated in up to the dose ranging from 0.1 to 94 dpa [39].

In conclusion, we would like to note another important consequence of the present results demonstrating thermally *activated nature of the dislocation unpinning* from radiation-induced defects. Temperature fluctuations are not the only mechanism of the activation of dislocation unpinning *under irradiation*, as have been argued by Dubinko et al [40]. The point is that irradiation produces strong lattice excitations (a.k.a. discrete breathers), recently proven to exist in bcc Fe [41, 42], and their interaction with dislocations can *activate unpinning* from structural defects. This may change mechanical properties of materials under reactor conditions as compared to the surveillance specimens in out-reactor tests after equivalent irradiation dose, as has been demonstrated experimentally by Grynik and Karasev [43]. So the temperature dependence of the pinning strength obtained in the present work should be modified to include also the dose rate effects (i.e. neutron flux, *F*, etc.), in order to predict evolution of mechanical properties of materials directly *under reactor irradiation conditions*: $\alpha_0(T) \rightarrow \alpha_0(T, F)$. This problem is beyond the scope of the present paper and will be addressed elsewhere.

### 4. Conclusions and outlook

1. In the commonly used Orowan stress ($\Delta\sigma_c = \alpha Gb/L$) the factor $\alpha$ should not be treated as a constant, since it depends not only on the type of obstacles but also on their number density that evolve with irradiation time/dose. The pinning strength factor $\alpha_0$ is defined, which characterizes the obstacle type more precisely.

2. Linear superposition of hardening due to DLs and precipitates has been shown to be adequate for evaluation of their strength factors in RPV irradiated in WWER-440 and WWER-1000.

3. The pinning strength of loops and precipitates at reactor operation temperatures are shown to be much lower than those measured for model alloys at room temperature as well as those evaluated by means of atomistic simulations in the temperature range of 300 to 600 K. This discrepancy is explained in the framework of a model of thermally activated dislocation motion, which takes into account the difference in temperature and strain rate employed in atomistic simulations and in mechanical testing.

4. Analytical expression for the obstacle pinning strength has been derived, which depends on the temperature and strain rate of mechanical testing and provides a link between MD simulations and experiment.

5. In order to predict evolution of mechanical properties of materials directly under reactor irradiation conditions one needs to take into account the pinning strength dependence on the neutron flux as well as on the irradiation temperature: $\alpha_0(T) \rightarrow \alpha_0(T, F)$.

**Acknowledgements**

This research is executed under the partial financial support of the European Union in the framework of the STCU Project (Grant #5497).